# Superconducting praseodymium superhydrides

Di Zhou[1], Dmitrii V. Semenok[2], Defang Duan[1], Hui Xie[1], Wuhao Chen[1], Xiaoli Huang[1,*], Xin Li[1], Bingbing Liu[1], Artem R. Oganov[2,3,*] and Tian Cui[1,*]

[1] State Key Laboratory of Superhard Materials, College of Physics, Jilin University, Changchun 130012, China

[2] Skolkovo Institute of Science and Technology, Skolkovo Innovation Center 143026, 3 Nobel Street, Moscow, Russia

[3] International Center for Materials Discovery, Northwestern Polytechnical University, Xi'an, 710072, China

*Corresponding authors: huangxiaoli@jlu.edu.cn, a.oganov@skoltech.ru and cuitian@jlu.edu.cn

**ABSTRACT**

Superhydrides have complex hydrogenic sublattices and are important prototypes for studying metallic hydrogen and high-temperature superconductors. Encouraged by the results on $LaH_{10}$, in consideration of the differences between La and Pr, Pr-H system is especially worth studying because of the magnetism and valence-band *f-electrons* in element Pr. Here we successfully synthesized praseodymium superhydrides ($PrH_9$) in laser-heated diamond anvil cells. Synchrotron X-ray diffraction (XRD) analysis demonstrated the presence of previously predicted $F\bar{4}3m$-$PrH_9$ and unexpected $P6_3/mmc$-$PrH_9$ phases. Moreover, $Fm\bar{3}m$-$PrH_3$, $P4/nmm$-$PrH_{3-\delta}$ and $Fm\bar{3}m$-$PrH_{1+x}$ were found below 52 GPa. $F\bar{4}3m$-$PrH_9$ and $P6_3/mmc$-$PrH_9$ were stable above 100 GPa in experiment. Experimental studies of electrical resistance in the $PrH_9$ sample showed the emergence of superconducting transition ($T_c$) below 9 K <span style="color:red">and a dependent $T_c$ on applied magnetic field.</span> Theoretical calculations indicate that magnetic order and electron-phonon interaction coexist in a very close range of pressures in the $PrH_9$ sample which may contribute to its low superconducting temperature $T_c$. Our results highlight the intimate connections among hydrogenic sublattices, density of states, magnetism and superconductivity in Pr-based superhydrides.

**Keywords:** high pressure, superhydrides, crystal structure, superconductivity

**SUMMARY**

Observations of high-temperature superconductivity in dense hydrogen-rich compounds have reinvigorated the field of high pressure and superconductivity ever since the striking discovery of $H_3S$ and $LaH_{10}$ with $T_c$ exceeding 200 K. Superhydrides with complex hydrogenic sublattices are important prototype systems to investigate metallization of hydrogen with potential high-temperature superconductivity. The present systematic experimental and computational study of the praseodymium hydrides reveals two phases of praseodymium surpehydride ($PrH_9$) with intriguing crystal structures and properties under pressure. Our experimental results show that superconductivity declines along the La-Ce-Pr series, while magnetism becomes more and more pronounced, indicating that lanthanide atoms play a more profound role in determining superconducting $T_c$.

# INTRODUCTION

The idea that hydrogen-rich compounds may be potential high-$T_c$ superconductors that can be traced back to 2004 (1), when chemical pre-compression of hydrogen due to bonding with other elements was proposed as an effective way to reduce the metallization pressure of hydrogen. Recent experimental results of $T_c$ exceeding 200 K in compressed $H_3S$ (2, 3, 4) and 250-260 K in $LaH_{10}$ system (5, 6, 7, 8) have indicated compressed hydrogen-rich compounds toward as potential room-temperature superconductors.

It is recognized that superconductivity in such hydrides owes its origin to electron-phonon coupling. According to BCS theory, from the expression of the critical temperature, three parameters determine $T_c$: the characteristic phonon frequency, electron-phonon coupling and Coulomb pseudo-potential (9). Recent theoretical studies have covered almost all binary hydrides, and found several metal superhydrides with extraordinary high-$T_c$ superconductivity, such as $CaH_6$ (10), $MgH_6$ (11), $YH_{6-10}$ (12, 13), $AcH_{10-16}$ (14), $ThH_{9-10}$ (15). Peng et al. (16) firstly studied all the candidate structures of rare earth superhydrides with H-rich cages at high pressure, and proposed that only several hydrides could be superconductors with $T_c > 77$ K. At the same time, superhydrides with $H_2$ units are recognized to have relatively low critical temperature, e.g. $LiH_6$ (17), $NaH_7$ (18), $Xe(H_2)_7$ (19) and $HI(H_2)_{13}$ (20). The question is why some superhydrides are high-$T_c$ superconductors, while others, with the same structure and stoichiometry, are not.

Continuing studies of lanthanide superhydrides, in this work we studied high-pressure behavior of the Pr-H system above 100 GPa. Chesnut and Vohra (21) studied the crystal structure of metallic Pr and determined the phase sequence above megabar pressure. Pr can readily absorb hydrogen at high temperature and form hydrides: face-centered cubic dihydride $PrH_2$ and hexagonal close-packed trihydride $PrH_3$ were found at ambient pressure. Subsequent filling of octahedral voids in the structure of dihydrides leads to non-stoichiometric $PrH_{2+x}$ composition which exhibit considerable variations of magnetic structures (22). Here, through high-pressure and high-temperature (HPHT) synthesis, two unexpected Pr superhydrides were obtained and studied. In particular, we investigated superconducting behavior of synthesized Pr superhydrides by electrical resistance measurements. Theoretical calculations are employed to deeply unearth the relationship among their magnetic properties, electronic band structures, phonon spectra and superconductivity. Comparison with already detailed studies of La and Ce superhydrides allows us to elucidate the great influence of metal atoms on superconductivity of superhydrides.

# RESULTS AND DISSCUSSION

## The stability and structures predicted by theoretical calculations

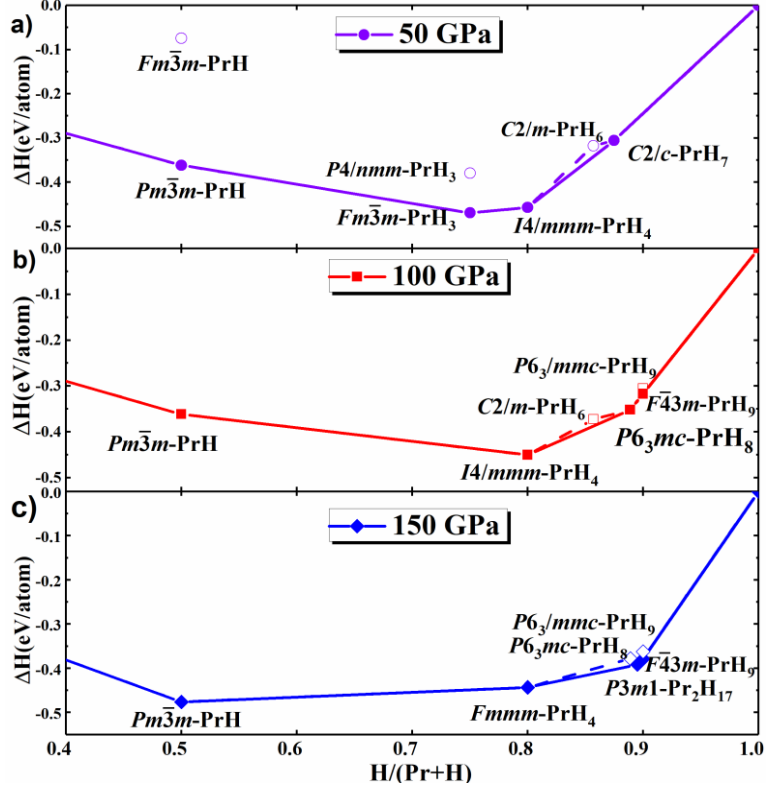

**Fig. 1. Calculated convex hulls for Pr-H system at various pressures.** Convex hulls for Pr-H system with spin-orbit coupling (SOC) and magnetic corrections at (**a**) 50 GPa, (**b**) 100 GPa and (**c**) 150 GPa.

Before describing the experimental results, we have compared our theoretical findings with the previous *ab initio* study (16), which is different from ours in a number of aspects. These differences are crucial for understanding our experimental results, and motivated us to further perform independent variable-composition searches for stable compounds in the Pr-H system at pressures of 50, 100 and 150 GPa using the USPEX (23, 24, 25) package and AIRSS (26) code (see Supplementary Materials Fig. S1). The current theoretical results performed by VASP (27, 28, 29) are also checked by an independent code CASTEP (30). The results of CASTEP can be found in Supplementary Materials Fig. S1. These two codes give the same results in principle. The only difference is the symmetry of $PrH_3$ that CASTEP gives $C2/m$-$PrH_3$ while VASP gives $Pm\bar{3}m$-$PrH_3$ without magnetism and spin-orbit coupling (SOC) effects.

Results of the structure search exhibit large differences depending on including or excluding magnetism and SOC effects, which can be seen in Fig. 1 and Supplementary Materials Fig. S1. However, previous calculations (16) didn´t include these effects. In agreement with previous results (16) our search gives $Pm\bar{3}m$ as the most stable symmetry for monohydride PrH and $Fm\bar{3}m$ for trihydride $PrH_3$, but, such important metastable phases for the further $P4/nmm$-$PrH_3$ (~70 meV/atom above the convex hull) and $Fm\bar{3}m$-PrH were not reported. Previous work indicated that superhydride $F\bar{4}3m$-$PrH_9$ is stable between 100 and 200 GPa, but no $P6_3/mmc$-$PrH_9$, which is about 19 meV/atom above the convex hull at 100 GPa. We also updated the convex hull and phase diagram of Pr-H system at 150 GPa.

# Synthesis of polyhydrides $Fm\bar{3}m$-PrH$_3$ and $P4/nmm$-PrH$_{3-\delta}$

To synthesize novel hydrides, we carried out several experiments by directly compressing Pr and hydrogen in the DACs. The diamond used in this experiment was coated with 150 nm alumina film by magnetron sputtering. The metallic Pr sample was loaded and sealed with a little pressure in the argon-protected glove box. After loading hydrogen into the cell, the sealed pressure was about 10 GPa, and selected XRD patterns are shown at various pressure (see Supplementary Materials Fig. S3c). Before laser heating, the diffraction pattern at 30 GPa included peaks from $Fm\bar{3}m$-PrH$_3$ in Fig. 2a, the structure of which can be viewed as cubic close packing of Pr atoms with all octahedral and tetrahedral voids filled by H atoms. (see Fig. 2c). After compression to 40 GPa, the sample was laser-heated to 1400 K. We found stronger signal from $Fm\bar{3}m$-PrH$_3$, while peaks from $Fm\bar{3}m$-PrH disappeared (see Supplementary Materials Fig. S3c). Upon further compression, the diamonds were broken.

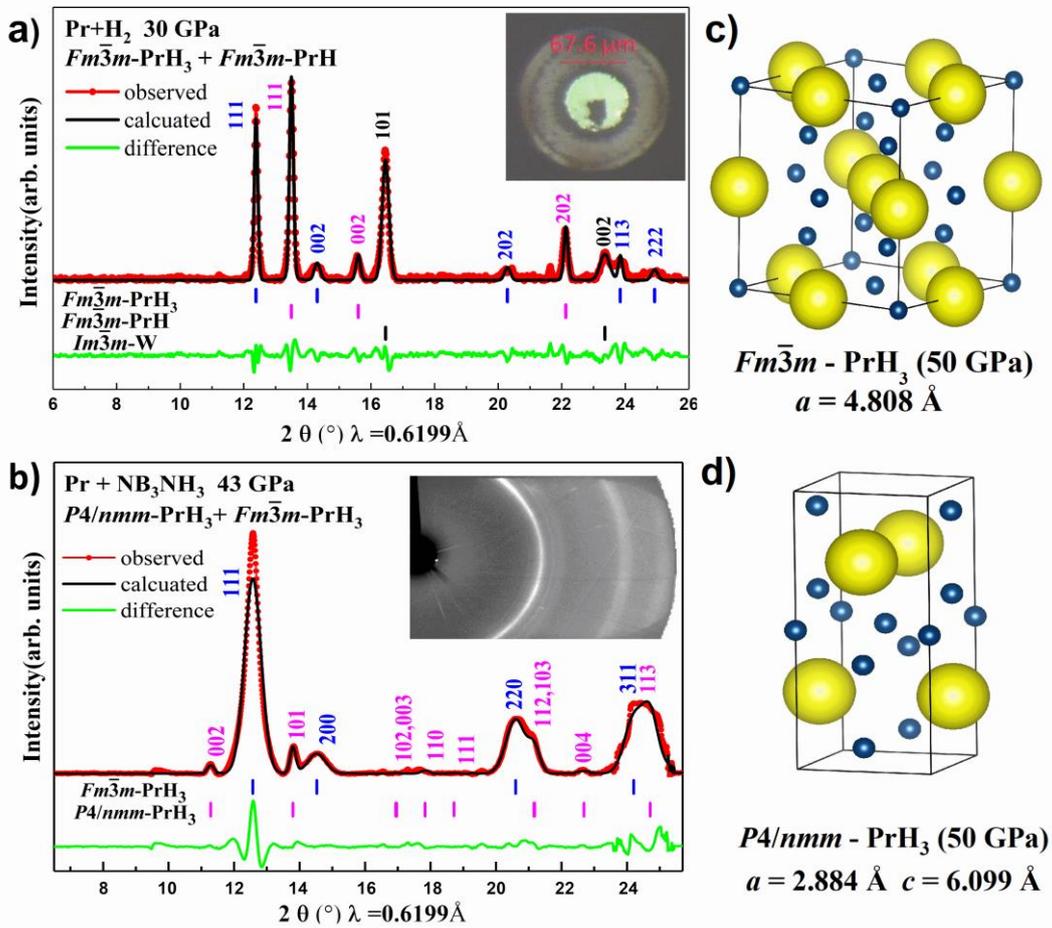

**Fig. 2. XRD patterns and crystal structures of PrH$_3$ at pressures.** (a) Refinement of the experimental XRD patterns obtained in Pr + H$_2$ cell by cold compression to 30 GPa. (b) Refinement of the XRD pattern by $Fm\bar{3}m$-PrH$_3$ and $P4/nmm$-PrH$_3$ after laser-heating at 43 GPa. Red line: experimental data; black line: model fit for the structure; green line: residues. Reliable parameters for the refinement are $R_p$ = 14.2 %, $R_{wp}$ = 24.5 %. Crystal structures of (c) $Fm\bar{3}m$-PrH$_3$ and (d) $P4/nmm$-PrH$_3$ phase at 50 GPa.

The experimental volumes of cubic PrH$_3$ are in good agreement with the predicted $Fm\bar{3}m$-PrH$_3$ structure in the pressure range 10-53 GPa (see Fig. 3b). The experimental equation of state (EoS) of this phase was fitted by the third-order Birch-Murnaghan equation of state (3$^{rd}$ B-M), which gave $V_0$ = 37.7 (3) Å$^3$, $K_0$ = 113 (2)

GPa, and $K_0' = 3.0$ (5). $Fm\bar{3}m$-PrH, proposed for explanation of the XRD pattern, is slightly non-stoichiometric from the EoS (Fig.3c), it seems more correct to define as $Fm\bar{3}m$-PrH$_{1+x}$ where $x = 0.08 - 0.13$.

It is well known that experimental studies of hydrides are greatly affected by the hydrogen permeability contributing to the failure of the high-pressure experiments. To minimize this problem, we synthesized the new hydrides by replacing of pure hydrogen with ammonia borane (AB), which is an excellent source of hydrogen (released during decomposition of AB). Several experiments were performed according to the reaction: Pr + NH$_3$BH$_3$→PrH$_x$ + $c$-BN through HPHT treatment (31, 32, 33). Fig. 2b shows the diffraction pattern after laser heating at 43 GPa. The reacted products are mainly dominated by $Fm\bar{3}m$-PrH$_3$ with a small quantity of tetragonal phase $P4/nmm$-PrH$_{3-\delta}$ ($0.05 \leq \delta \leq 0.15$) with smaller unit cell volume. At 50 GPa in the $P4/nmm$-PrH$_3$ structure, each Pr atom is linked with 9 H atoms with 2.09 Å $\leq d$(Pr-H) $\leq$ 2.17 Å. Experimental cell parameters of discovered compounds are shown in Supplementary Materials Table S3.

## Synthesis of $F\bar{4}3m$-PrH$_9$ and $P6_3/mmc$-PrH$_9$

To obtain higher hydrides of Pr, we conducted further experiments at pressures above 100 GPa. To overcome problems with hydrogen permeation, we also used NH$_3$BH$_3$ (AB) as the source of hydrogen, which proved to be effective for synthesis of superhydrides at megabar pressures (7, 34). The original sample containing Pr with AB was laser-heated to 1650 K at 115 GPa. Measurements after laser-heating did not show any changes in pressure, and Raman signal of H$_2$ was detected at 4147 cm$^{-1}$, indicating the generation of hydrogen. Fig. 3a shows the XRD pattern with the presence of two praseodymium superhydrides $F\bar{4}3m$-PrH$_9$ and $P6_3/mmc$-PrH$_9$. Experimental lattice parameters at 120 GPa are $a = 4.967$ (1) Å, $V = 122.52$ (9) Å$^3$ for $F\bar{4}3m$-PrH$_9$ and $a = 3.588$ (1) Å, $c = 5.458$ (4) Å, $V = 60.84$ (9) Å$^3$ for $P6_3/mmc$-PrH$_9$. This sample was compressed to 130 GPa and then gradually decompressed down to the lowest pressure 105 GPa to determine its experimental equation of state (Fig. 3b, Table S4). Both EoS of PrH$_9$ are very close to the calculated curve of Pr+9H and correspond well with the calculated values. After decompression, the cell dropped down to 53 GPa, and recorded XRD pattern demonstrates the presence of two lower hydride phases: $Fm\bar{3}m$-PrH$_3$ with experimental parameters of $a = 4.832$ (1) Å at 50 GPa, and $P4/nmm$-PrH$_{3-\delta}$ with $a = 2.801$ (1) Å, $c = 6.280$ (2) Å at 50 GPa, which is consistent with the low pressure results.

Both structures have almost the same volume and energy on convex hull at studied pressure range (Fig. 1b-1c). It is interesting to note that stability of $F\bar{4}3m$-PrH$_9$ was previously predicted (16), while its coexistence with metastable $P6_3/mmc$-PrH$_9$ is a surprise. According to our theoretical calculations, the enthalpy difference between $P6_3/mmc$-PrH$_9$ and $F\bar{4}3m$-PrH$_9$ is about 19 meV/atom which is near the limit of DFT accuracy. According to recent studies (35, 36), 20% of experimentally synthesized materials are metastable, some of which even have high positive.

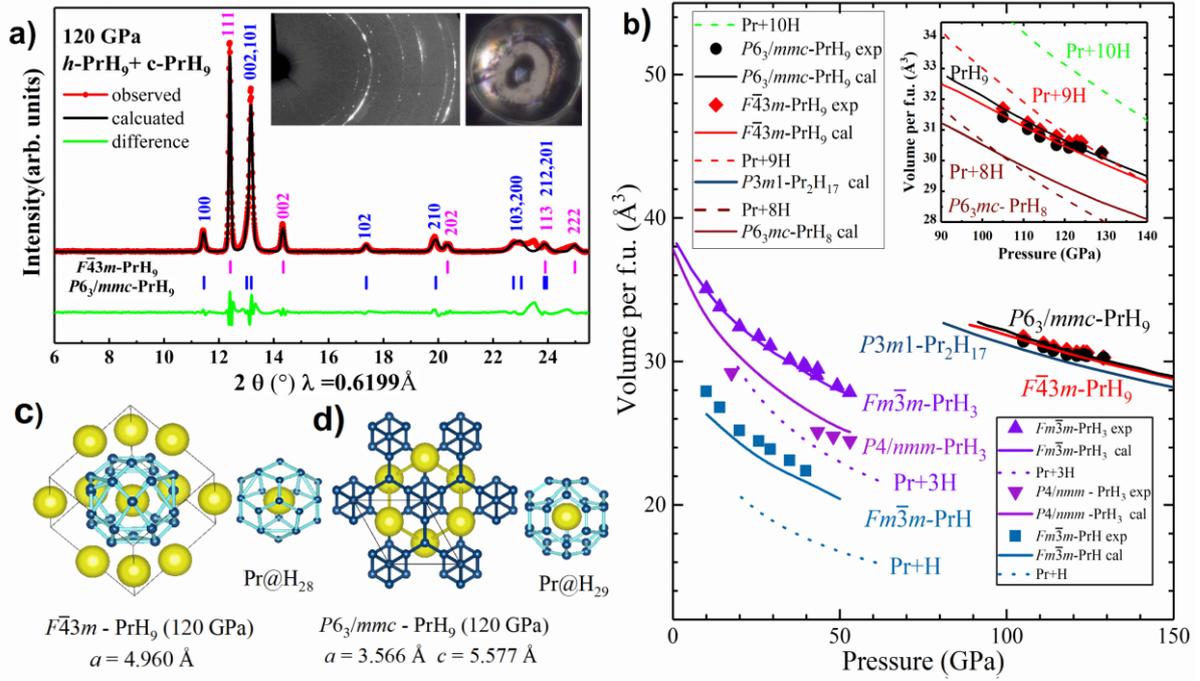

**Fig. 3. Refinement of the experimental XRD pattern, pressure-volume data and crystal structure of PrH$_9$.** (**a**) Refinement of the XRD pattern by $F\bar{4}3m$-PrH$_9$ and $P6_3/mmc$-PrH$_9$. Red line: experimental data; black line: model fit for the structure; green line: residues. Reliable parameters are as $R_p$ = 12.4%, $R_{wp}$ = 22.0%. (**b**) EoS of the synthesized Pr-H phases; theoretical results include magnetism and SOC effects. Inset: the distinction among PrH$_8$, PrH$_{10}$ and PrH$_9$ phases. Crystal structures of (**c**) $F\bar{4}3m$-PrH$_9$ with H$_{28}$ cages and (**d**) $P6_3/mmc$-PrH$_9$ with H$_{29}$ cages.

## Properties of $F\bar{4}3m$-PrH$_9$ and $P6_3/mmc$-PrH$_9$

We performed a series of experiments to investigate superconductivity of PrH$_9$ *via* measurements of electrical resistance $R$(T) in the range of 1.6 to 300 K at pressures from 100 GPa up to 150 GPa. The XRD pattern of the prepared sample at 126 GPa deposited with four electrodes, shows presence of both $F\bar{4}3m$-PrH$_9$ and $P6_3/mmc$-PrH$_9$ phases (Fig. 4c). Possible superconducting transitions were detected with the resistance drop below 9 K, so we proposed that the superconducting transition temperature is below 9 K, far below LaH$_{10}$ of the same main group. The superconducting resistance drop $R$(T) is also dependent on applied magnetic field, further proving this is a superconducting transition (see Fig. 4d). Another run of experiments confirmed the existence of the pronounced superconducting resistance drop in PrH$_9$ below 9 K (see Fig. 4e and Supplementary Materials Fig. S9). The complexity of the experiments prevented us from accurately determining the pressure dependence of superconducting $T_c$. We did not observe zero resistance of the superhydrides samples because of their complex geometries, and the samples were clearly mixed phase, possible with varying hydrogen stoichiometry. The same phenomena of incompletely dropping to zero in resistance have also been reported in the measurement of superconducting resistance of boron (37) and iron (38) at high pressure.

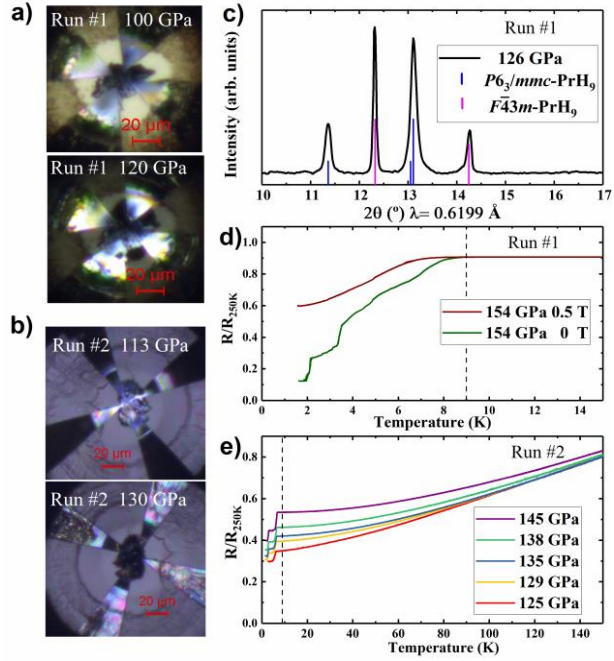

**Fig. 4. Electrical resistance measurements of PrH$_9$.** (a) The sample inside the diamond anvil cell connected with four electrodes before and after laser heating for sample 1. (b) The photos of sample 2 from different sides of cell after heating. (c) XRD pattern proves the cubic and hexagonal PrH$_9$ were synthesized in the sample at around 120 GPa from a mixture of Pr and AB. (d) Resistance steps of sample 1 at different magnetic field. (e) Resistance steps of sample 2 at different pressure.

Further theoretical calculations try to understand why both $F\bar{4}3m$-PrH$_9$ and $P6_3/mmc$-PrH$_9$ possess such low $T_c$. As shown in Fig. 3c-3d, both structures have clathrate structures, which are also found in other rare earth hydrides. Calculations of the electron localization function (ELF) reveal weak covalent H-H interactions. In $F\bar{4}3m$-PrH$_9$ structure, the nearest H-H distance is 1.135 Å at 120 GPa, which is a bit longer than the known shortest H-H distance in $P6_3/mmc$-CeH$_9$ (~ 1.1 Å) (39), but shorter than $Fm\bar{3}m$-LaH$_{10}$ (5). At the same time, $P6_3/mmc$-PrH$_9$ with Pr@H$_{29}$ cages has the nearest H-H distance of ~1.170 Å (at 120 GPa), which is longer than $d_{\min}$(H-H) in atomic hydrogen as well as in CeH$_9$ at the same pressure (16) (for details see Fig. S4).

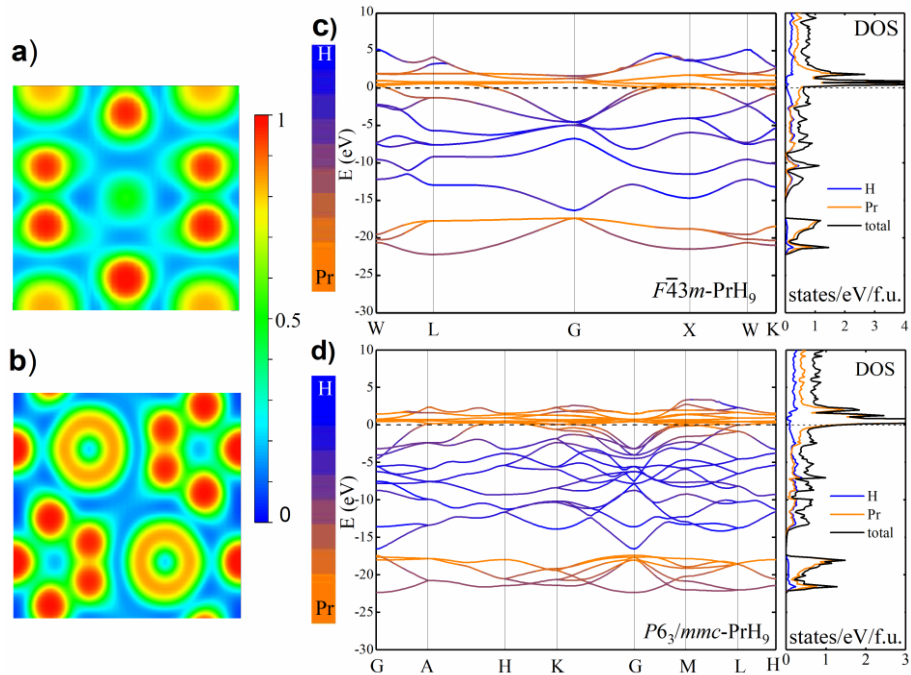

**Fig. 5. Electronic properties of PrH$_9$.** Electron localization function (ELF) of (a) $F\bar{4}3m$-PrH$_9$ and (b) $P6_3/mmc$-PrH$_9$. Calculated densities of electron states (DOS) and band structure in (c) $F\bar{4}3m$-PrH$_9$ and (d) $P6_3/mmc$-PrH$_9$ at 150 GPa.

DOS ($E_F$) is mostly due to *f-electrons* of Pr and has very high value in both cases.

Calculations demonstrate that both PrH$_9$ structures are dynamically stable (Fig. S6) and exhibit metallic properties (Fig. 5). But only 6 to 9 % of the total DOS comes from the hydrogen atoms, the rest being due to *f-electrons* of Pr. Relatively high values of the density of states above 3-4 eV$^{-1}$f.u.$^{-1}$ at or near ($\pm$ 1 eV) the Fermi level caused by a series of Van Hove singularities, make it impossible to use constant DOS approximation when calculating parameters of the superconducting state in PrH$_9$ (40). Low contribution of hydrogen to DOS is associated with weak electron-phonon coupling (EPC) at 150 GPa, resulting in low superconducting $T_c$. EPC calculations for both PrH$_9$ with the selected PP give the estimated $T_c$ of 0.8 K for cubic PrH$_9$ and 8.4 K for hexagonal PrH$_9$ at 120 GPa with $\mu^* = 0.1$, which is good agreement with experiments (see Supplementary materials, Fig. S10 - S12).

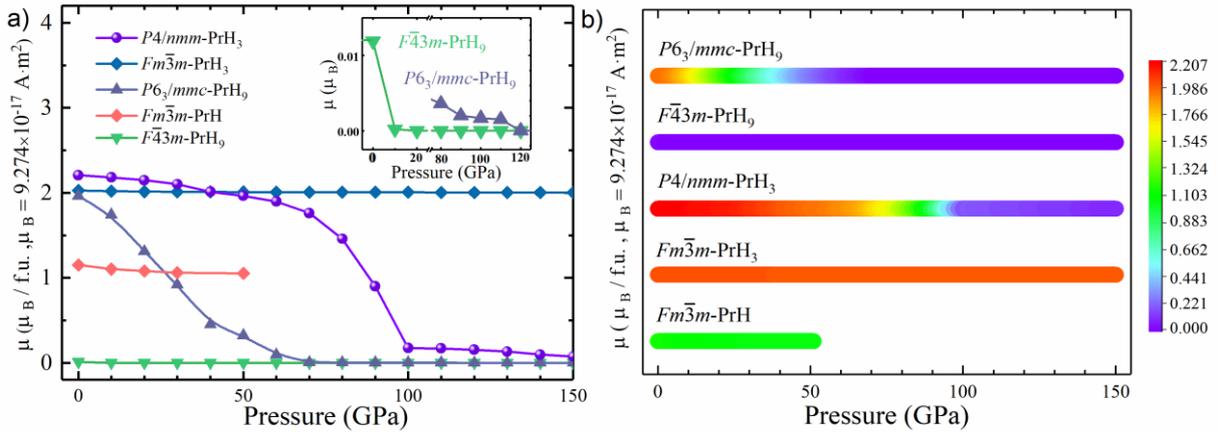

**Fig. 6. Magnetism of Pr hydrides at pressures up to 150 GPa.** (**a**) Magnetic moments of Pr-H system at high pressure and (**b**) Magnetic map of Pr-H system as a function of pressure.

We summarized magnetic properties for all studied praseodymium hydrides at pressure range 0-150 GPa in Fig. 6. We find that all Pr-H compounds are magnetic: $Fm\bar{3}m$-PrH$_3$ and $Fm\bar{3}m$-PrH possess strong magnetism and retain almost constant magnetic moments at high pressures, while tetragonal PrH$_3$ and both phases of PrH$_9$ lose magnetism under pressure. $P6_3/mmc$-PrH$_9$ loses magnetism at 120 GPa, while $F\bar{4}3m$-PrH$_9$ retains a very low magnetic moment. Magnetic order and electron-phonon interaction coexist in a very close range of pressures in praseodymium hydrides which may have an effect on the low superconducting transition temperature $T_c$.

## CONCLUSIONS

By the efficient means of *in situ* decomposition reaction of NH$_3$BH$_3$ under HPHT conditions previously used for synthesis of lanthanum superhydrides, we synthesized two novel metallic superhydrides $F\bar{4}3m$-PrH$_9$ and $P6_3/mmc$-PrH$_9$, two trihydrides $Fm\bar{3}m$-PrH$_3$ and $P4/nmm$-PrH$_{3-\delta}$, and one monohydride $Fm\bar{3}m$-PrH$_{1+x}$ in the pressure range 0 – 130 GPa. For most synthesized phases the equations of state and unit cell parameters are in good agreement with our DFT calculations. Resistance measurements of praseodymium hydrides indicated possible superconducting transitions in both PrH$_9$ were below 9 K, which is in agreement with theoretical calculations: 8.4 K for hexagonal and 0.8 K for cubic PrH$_9$ at 120 GPa. Magnetic order and electron-phonon interaction coexist in a very close range of pressures in praseodymium hydrides which may have an effect on the low superconducting transition temperature. Present results on Pr superhydrides show that superconductivity declines along the La-Ce-Pr series, while magnetism becomes more and more pronounced. Metallic atoms are not just donors of the electrons to the "metallic hydrogen" sublattice, but play a more profound role in determining superconducting $T_c$.

## METHODS

**Experimental method:** The praseodymium powder samples were purchased from Alfa Aesar with a purity of 99.99%. Molybdenum electrodes were sputtered onto the surface of one diamond anvils in the van der Pauw four-probe geometry. A four-probe measurement scheme was essential to separate the sample signal from the parasitic resistance of the current leads. We prepared an isolated layer from cubic boron nitride (or a mixture of epoxy and $CaF_2$). We performed laser heating of three diamond anvil cells (100 μm and 150 μm culets) loaded with metallic Pr sample and ammonia borne in the argon-protected glove box. The diamonds used for electrical DACs had a culet with a diameter of 100 μm. Thickness of the tungsten gasket was 20±2 μm. Heating was carried out by pulses of infrared laser with wavelength 1 μm (Nd:YAG) and temperature measurements were carried out by the MAR 345 detector. Applied pressure was measured by the edge position of diamond Raman signal (41). X-ray diffraction patterns studied in diamond anvil cells samples were recorded on the BL15U1 synchrotron beamline (42) at Shanghai Synchrotron Research Facility (SSRF, China) with the use of a focused (5×12 μm) monochromatic. Additional syntheses with electrodes were carried out at the 4W2 High-Pressure Station of Beijing Synchrotron Radiation Facility (BSRF, China). The beam size was about 32×12 μm. both facilities are with the incident X-ray beam (20 keV, 0.6199 Å) and a Mar165 CCD two-dimensional as the detector. The experimental X-ray diffraction images were analyzed and integrated using the Dioptas software package (43). The full profile analysis of the diffraction patterns, as well as the calculation of the unit cell parameters, was performed in the Materials studio (44) and JANA2006 program (45) by the Le Bail method (46).

**Theoretical calculations:** We have carried out variable-composition searches for stable compounds in the Pr-H system at pressures of 50, 100 and 150 GPa using the USPEX (23, 24, 25) package and AIRSS (26) code coupled with the Cambridge serial total energy package (CASTEP) plane-wave code (30) and on the fly pseudopotentials (47). The first generation of USPEX search (120 structures) was created using a random symmetric generator, while all subsequent generations (100 structures) contained 20% random structures and 80% created using heredity, softmutation, and transmutation operators.

We calculated the equation of states (EoS) for PrH, both $PrH_3$ and two $PrH_9$ phases. In order to calculate the equations of state, we performed structure relaxations of phases at various pressures using density functional theory (DFT) (48, 49) within the generalized gradient approximation (Perdew-Burke-Ernzerhof functional) (50, 51) and the projector-augmented wave method (52, 53) as implemented in the VASP code (27, 28, 29). Plane wave kinetic energy cutoff was set to 1000 eV and the Brillouin zone was sampled using Γ-centered $k$-points meshes with resolution $2\pi \times 0.05$ Å$^{-1}$. Obtained dependences of the unit cell volume on pressure were fitted by three-order Birch-Murnaghan equation (54) to determine the main parameters of the EoS, namely $V_0$, $K_0$ and $K'$, where $V_0$ is equilibrium volume, $K_0$ is bulk modulus and $K'$ is derivative of bulk modulus with respect to pressure using the EoSfit7 software (55). We also calculated phonon densities of states for studied materials using finite displacement method (VASP (56) and PHONOPY (57)).

Calculations of phonons, electron-phonon coupling and superconducting $T_c$ were carried out with QUANTUM ESPRESSO (QE) package (58) using density-functional perturbation theory (59), employing plane-wave pseudopotential method and local-density approximation exchange-correlation functional (60). Norm-conserving potentials for H ($1s^1$) and Pr ($5s^2 5p^6 4f^3 6s^2$) were used with a kinetic energy cutoff of 90 Ry. In our *ab initio* calculations of the electron-phonon coupling (EPC) parameter $\lambda$, the first Brillouin zone was sampled using a 6×6×6 q-points mesh with a denser 24×24×24 $k$-points mesh for $F\bar{4}3m$-PrH$_9$, and a 3×3×2 q-points mesh with a denser 15×15×10 $k$-points mesh for $P6_3/mmc$-PrH$_9$ (with Gaussian smearing and $\sigma = 0.035$ Ry which approximates the zero-width limits in the calculation of $\lambda$). Critical temperature $T_c$ was calculated from the Allen-Dynes modified McMillan formula (61): $Tc = \frac{\omega_{log}}{1.2} exp\left[-\frac{1.04(1+\lambda)}{\lambda - \mu^*(1+0.62\lambda)}\right]$, with $\omega_{log} = exp[\frac{2}{\lambda} \int ln(\omega) \frac{\alpha^2 F(\omega)}{\omega} d\omega]$ and $\lambda = 2 \int \frac{\alpha^2 F(\omega)}{\omega} d\omega$, where $\mu^*, \alpha^2 F(\omega)$ and $\lambda$ are Coulomb parameter, the electron-phonon spectral function and the EPC parameter.

## SUPPLEMENTARY MATERAILS

**Table S1** Crystal structure of predicted Pr-H phases.
**Table S2** Experimental parameters of DACs.
**Table S3** Experimental cell parameters and volumes of lower praseodymium hydrides along with calculated cell volumes ($V_{DFT}$).
**Table S4** Experimental cell parameters and volumes of two praseodymium superhydrides along with calculated cell volumes ($V_{DFT}$).
**Table S5** Reference EoS parameters of pure elemental Pr in Birch–Murnaghan equation.
**Table S6** Calculated EoS parameters of 3$^{rd}$ Birch–Murnaghan equation for all studied Pr-H phases.

**Fig. S1** Calculated convex hulls for Pr-H system at various pressures.

**Fig. S2** Convex hulls without and with ZPE correction and crystal structure of discovered praseodymium hydrides at 120 GPa.

**Fig. S3** Experimental XRD patterns dependence of pressure in the range of 0-130 GPa.

**Fig. S4** (**a**) Comparison of the pressure dependence of the nearest H-H distances for two PrH$_9$, LaH$_{10}$, CeH$_9$, and atomic H. (**b**) Nearest Pr-H distances as a function of pressure calculated from experimental cell parameters.

**Fig. S5** Raman spectra of Z1 cell under decompression.

**Fig. S6** Calculated phonon density of states and band structure for $P6_3/mmc$-PrH$_9$ (**a**) and $F\bar{4}3m$-PrH$_9$ (**b**) at 120 GPa.

**Fig. S7** Calculated phonon density of states and band structure for $P6_3mc$-PrH$_8$ (**a**) at 120 GPa and $P4/nmm$-PrH$_3$ at 50 GPa (**b**).

**Fig. S8** Electron density of states for cubic PrH$_3$ and $P4/nmm$-PrH$_3$ at 30 GPa.

**Fig. S9** Enlarged figure of electrical resistance measurements of PrH$_9$ in sample 2.

**Fig. S10** Calculated superconducting parameters of $F\bar{4}3m$-PrH$_9$ at 120 GPa as a function of electronic smearing σ and the pseudopotential.

**Fig. S11** Eliashberg spectral functions, the electron-phonon integral $λ(ω)$ and critical transition temperature $T_c(ω)$ calculated at 120 GPa for cubic PrH$_9$ with $σ = 0.035$ Ry.

**Fig. S12** A series of Eliashberg spectral functions calculated at 120 GPa for hexagonal PrH$_9$.

# Acknowledgments

**Funding:** This work was supported by the National Key R&D Program of China (Grant No. 2018YFA0305900), National Natural Science Foundation of China (Grant Nos. 51572108, 51632002, 11974133, 11674122, 11574112, 11474127, and 11634004), National Key Research and Development Program of China (Grant No. 2016YFB0201204), Program for Changjiang Scholars and Innovative Research

Team in University (Grant No. IRT_15R23), and National Fund for Fostering Talents of Basic Science (Grant No. J1103202). A. R. O. thanks Russian Science Foundation (Grant No. 19-72-30043). D. V. S. thanks Russian Foundation for Basic Research, grant No. 19-03-00100 A. The authors express their gratitude to the staffs of BL15U and 4W2 stations of Shanghai and Beijing Synchrotron Radiation Facilities.

**Author Contributions:**

D. Z., D. V. S. and D. D. contributed equally to this work.

X. H., A. R. O. and T. C. conceived this project. D. Z., D. V. S., and X. H. performed the experiment, D. V. S., D. D. A. R. O. and T. C. prepared theoretical calculations and analysis. X. H., D. Z., D. V. S., A. R. O. and T. C. wrote and revised the paper. All authors discussed the results and offered the useful discussions.

**Competing interests:** The authors declare no competing interests.

**Data and materials availability:** All data needed to evaluate the conclusions in the paper are present in the paper and/or the Supplementary Materials. Additional data related to this paper may be requested from the authors.

# Supplementary Materials
# for
## Superconducting praseodymium superhydrides

Di Zhou[1], Dmitrii V. Semenok[2], Defang Duan[1], Hui Xie[1], Wuhao Chen[1], Xiaoli Huang[1,*], Xin Li[1], Bingbing Liu[1], Artem R. Oganov[2,3,*] and Tian Cui[1,*]

[1] State Key Laboratory of Superhard Materials, College of Physics, Jilin University, Changchun 130012, China
[2] Skolkovo Institute of Science and Technology, Skolkovo Innovation Center 143026, 3 Nobel Street, Moscow, Russia
[3] International Center for Materials Discovery, Northwestern Polytechnical University, Xi'an, 710072, China

**Author Contributions:**
E. Z., D. V. S. and D. D. contributed equally to this work.
*Corresponding authors: huangxiaoli@jlu.edu.cn, a.oganov@skoltech.ru and cuitian@jlu.edu.cn

# CONTENT

# Structural information

**Table S1.** Crystal structure of predicted Pr-H phases.

| Phase | Pressure, GPa | Lattice parameters | Coordinates | | | |
|---|---|---|---|---|---|---|
| $F\bar{4}3m$-PrH$_9$ | 120 | $a$ = 4.960 Å<br>$\alpha = \beta = \gamma = 90°$ | Pr (4b) | 0.50000 | 0.50000 | 0.50000 |
| | | | H (16e) | 0.13659 | 0.13659 | 0.13659 |
| | | | H (16e) | -0.11826 | -0.11826 | -0.11826 |
| | | | H (4d) | 0.75000 | 0.75000 | 0.75000 |
| $P6_3/mmc$-PrH$_9$ | 120 | $a = b$ = 3.566 Å<br>$c$ = 5.577 Å<br>$\alpha = \beta = 90°$<br>$\gamma = 120°$ | Pr (2d) | 0.33333 | 0.66667 | 0.75000 |
| | | | H (12k) | 0.15844 | 0.31687 | 0.05586 |
| | | | H (4f) | 0.33333 | 0.66667 | 0.34188 |
| | | | H (2b) | 0.00000 | 0.00000 | 0.25000 |
| $Fm\bar{3}m$-PrH$_3$ | 50 | $a$ = 4.808 Å<br>$\alpha = \beta = \gamma = 90°$ | Pr (4b) | 0.50000 | 0.50000 | 0.50000 |
| | | | H (4a) | 0.00000 | 0.00000 | 0.00000 |
| | | | H (8c) | 0.25000 | 0.25000 | 0.25000 |
| $P4/nmm$-PrH$_3$ | 50 | $a = b$ = 2.884 Å<br>$c$ = 6.099 Å<br>$\alpha = \beta = \gamma = 90°$ | Pr (2c) | -0.50000 | 0.00000 | 0.74831 |
| | | | H (2c) | -0.50000 | 0.00000 | 0.37384 |
| | | | H (2b) | -0.50000 | 0.50000 | 0.50000 |
| | | | H (2c) | 0.00000 | 0.50000 | -0.09063 |
| $Fm\bar{3}m$-PrH | 50 | $a$ = 4.338 Å<br>$\alpha = \beta = \gamma = 90°$ | Pr (4b) | 0.50000 | 0.50000 | 0.50000 |
| | | | H (4a) | 0.00000 | 0.50000 | 0.50000 |

**Table S2.** Experimental parameters of DACs.

| #cell | Pressure of heating, GPa | Gasket | Sample size, μm | Composition/load |
|---|---|---|---|---|
| Z1 | 110 | W | 17/30 | Pr/BH$_3$NH$_3$ |
| Z2 | 40 | W | 25 | Pr/H$_2$ |
| Z3 (sample 1) | 100/120 | W | 35/50 | Pr/BH$_3$NH$_3$ |
| Z4 (sample 2) | 114 | W | 50 | Pr/BH$_3$NH$_3$ |

**Table S3.** Experimental cell parameters and volumes of lower praseodymium hydrides along with calculated cell volumes ($V_{DFT}$).

| P (GPa) | $Fm\bar{3}m$-PrH$_3$ (Z=4) | | | $Fm\bar{3}m$-PrH (Z=4) | | | P (GPa) | $P4/nmm$-PrH$_3$ (Z=2) | | | |
|---|---|---|---|---|---|---|---|---|---|---|---|
| | $a$ (Å) | $V$ (Å$^3$) | $V_{DFT}$ (Å$^3$) | $a$ (Å) | $V$ (Å$^3$) | $V_{DFT}$ (Å$^3$) | | $a$ (Å) | $c$ (Å) | $V$ (Å$^3$) | $V_{DFT}$ (Å$^3$) |
| 10 | 5.196(0) | 140.31(2) | 140.11 | 4.815(1) | 111.63(6) | 105.29 | 53 | 2.795(2) | 6.264(2) | 48.94(9) | 50.18 |
| 14 | 5.133(0) | 135.25(1) | 135.38 | 4.750(1) | 107.17(6) | 101.83 | 48 | 2.807(1) | 6.291(2) | 49.55(5) | 51.18 |
| 20 | 5.061(1) | 129.69(7) | 129.92 | 4.653(1) | 100.73(6) | 97.412 | 43 | 2.819(2) | 6.318(2) | 50.21(9) | 52.27 |
| 26 | 5.026(1) | 127.02(8) | 125.41 | 4.607(0) | 97.81(3) | 93.35 | 17 | 2.965(2) | 6.645(3) | 58.41(8) | 61.93 |
| 29 | 4.991(1) | 124.39(8) | 123.17 | 4.571(0) | 95.52(1) | 91.19 | | | | | |
| 35 | 4.938(0) | 120.42(2) | 119.78 | 4.520(0) | 92.37(2) | 88.52 | | | | | |
| 39 | 4.909(1) | 118.34(6) | 117.00 | 4.473(1) | 89.48(5) | 86.40 | | | | | |
| 43 | 4.878(0) | 116.08(1) | 115.22 | | | | | | | | |
| 49 | 4.837(1) | 113.22(7) | 112.15 | | | | | | | | |
| 53 | 4.811(0) | 111.38(4) | 110.70 | | | | | | | | |

**Table S4.** Experimental cell parameters and volumes of two praseodymium superhydrides along with calculated cell volumes ($V_{DFT}$).

| P (GPa) | $P6_3/mmc$-PrH$_9$ (Z=2) | | | | $F\bar{4}3m$-PrH$_9$ (Z=4) | | |
|---|---|---|---|---|---|---|---|
| | a (Å) | c (Å) | V (Å$^3$) | $V_{DFT}$ (Å$^3$) | a (Å) | V (Å$^3$) | $V_{DFT}$ (Å$^3$) |
| 105 | 3.635(1) | 5.492(3) | 62.83(8) | 63.33 | 5.024(1) | 126.77(8) | 126.00 |
| 111 | 3.615(1) | 5.479(0) | 62.05(4) | 62.57 | 4.999(1) | 124.93(5) | 124.14 |
| 114 | 3.604(1) | 5.473(3) | 61.55(8) | 62.01 | 4.987(1) | 123.99(7) | 123.45 |
| 118 | 3.595 (1) | 5.468(4) | 61.01(9) | 61.57 | 4.976(1) | 123.18(9) | 122.40 |
| 121 | 3.588(1) | 5.458 (4) | 60.84(9) | 61.19 | 4.965(1) | 122.42(5) | 121.63 |
| 123 | 3.590(1) | 5.461(3) | 60.94(6) | 60.93 | 4.967(1) | 122.52(9) | 121.12 |
| 124 | 3.587 (1) | 5.458(4) | 60.83(9) | 60.76 | 4.965(1) | 122.41(5) | 120.96 |
| 129 | 3.580 (1) | 5.441(4) | 60.49(9) | 60.23 | 4.947(1) | 121.07(7) | 123.63 |

**Table S5.** Reference EoS parameters of pure elemental Pr in the modified universal equation of state (MUEOS) as below:

$$\ln H = \ln B_0 + \eta(1-x) + \beta(1-x)^2 \qquad \text{S1}$$

where $x^3 = \frac{V}{V_0}$ is the volume compression, $\eta = 1.5(B_0' - 1)$, and $H = Px^2/[3(1-x)]$. $V_0$, $B_0$, $B_0'$ are the atomic volume, isothermal bulk modulus and the first pressure derivative of the bulk modulus at the ambient pressure.

| | $dhcp$-Pr | $C2/m$-Pr | $Cmcm$-Pr |
|---|---|---|---|
| $V_0$ (GPa) | 34.538 | 34.538 | 29.012 |
| $K_0$ (GPa) | 27.16(2) | 27.16(4) | 24.73(1) |
| $K'$ | 0.096(4) | 1.424(5) | 1.933(3) |
| $\beta$ | 39.88(2) | 25.25(6) | 22.35(1) |

**Table S6.** Calculated EoS parameters of 3$^{rd}$ Birch–Murnaghan equation for all studied Pr-H phases. For both PrH$_9$ $V_0$ corresponds to 100 GPa, for both PrH$_3$ and PrH correspond to 0 GPa.

| | $F\bar{4}3m$-PrH$_9$ | $P6_3/mmc$-PrH$_9$ | $Fm\bar{3}m$-PrH$_3$ | $P4/nmm$-PrH$_3$ | $Fm\bar{3}m$-PrH |
|---|---|---|---|---|---|
| $V_0$ (GPa) | 31.91 | 31.68 | 37.73 | 32.12 | 32.07 |
| $K_0$ (GPa) | 475.88 | 398.19 | 113.19 | 150.20 | 53.04 |
| $K'$ | 5.507 | 19.74 | 3.047 | 2.09 | 4.122 |

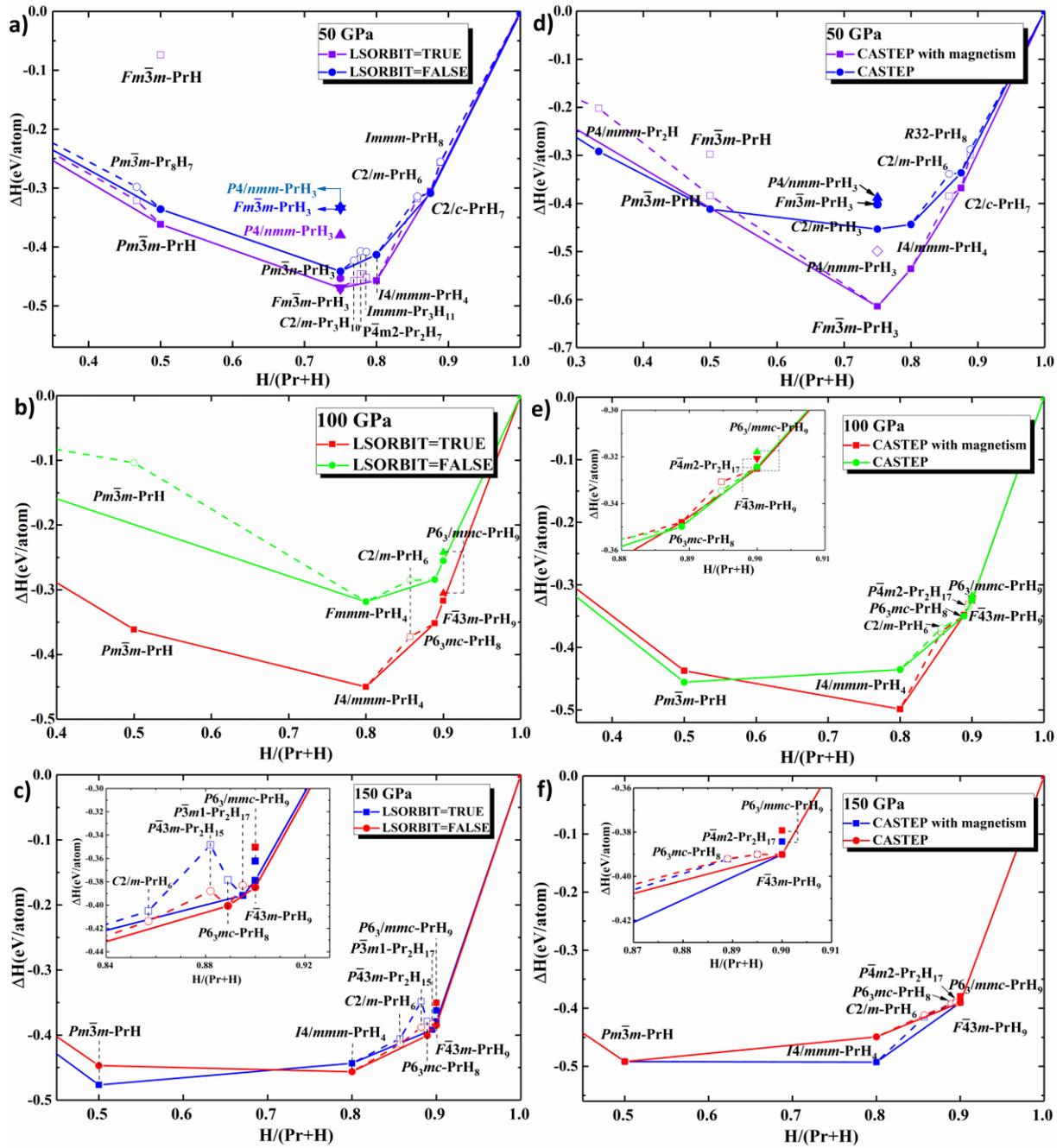

**Fig. S1. Calculated convex hulls for Pr-H system at various pressures.** Convex hulls for Pr-H system with and without spin-orbital and magnetic corrections calculated in USPEX at (**a**) 50 GPa (**b**) 100 GPa and (**c**) 150 GPa. The convex hulls of Pr-H systems calculated in AIRSS at (**d**) 50 GPa (**e**) 100 GPa and (**f**) 150 GPa. Inset shows the stability of hexagonal $PrH_8$ and cubic $PrH_9$.

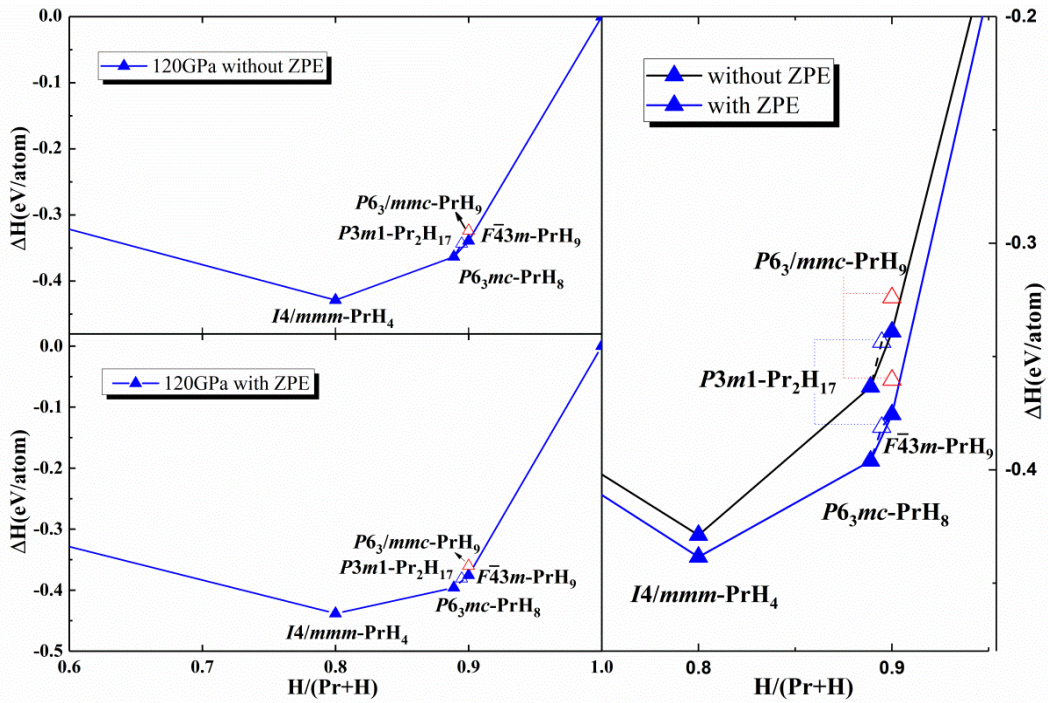

**Fig. S2.** Convex hulls without and with ZPE correction of discovered praseodymium hydrides at 120 GPa.

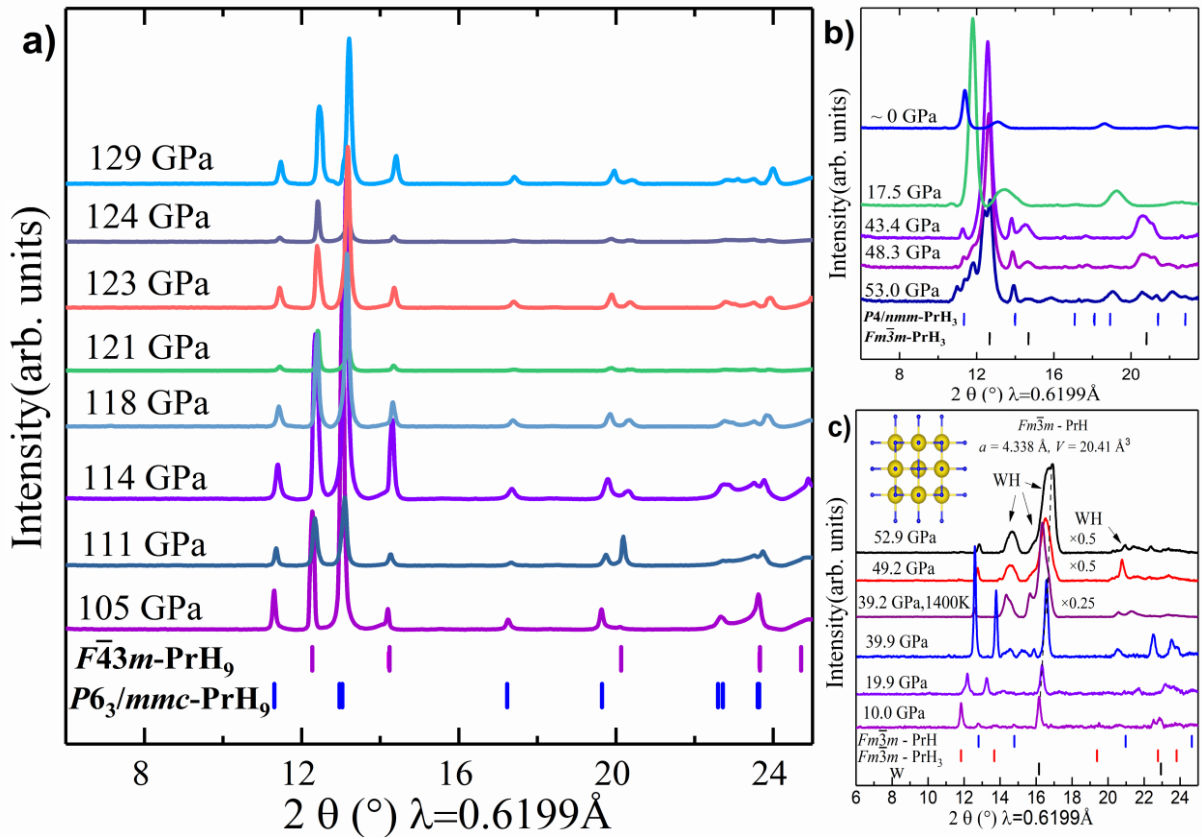

**Fig. S3.** Experimental XRD patterns dependence of pressure in the range of 0-130 GPa. **(a)** Experimental XRD patterns for the mixture of PrH$_9$ in the range of 129-105 GPa. **(b)** The XRD patterns upon decompression. **(c)** The selected XRD patterns during the compressing process of Pr with pure H$_2$. $Fm\bar{3}m$-PrH$_3$ and $Fm\bar{3}m$-PrH were synthesized at low pressure, after heating the peaks of $Fm\bar{3}m$-PrH disappeared.

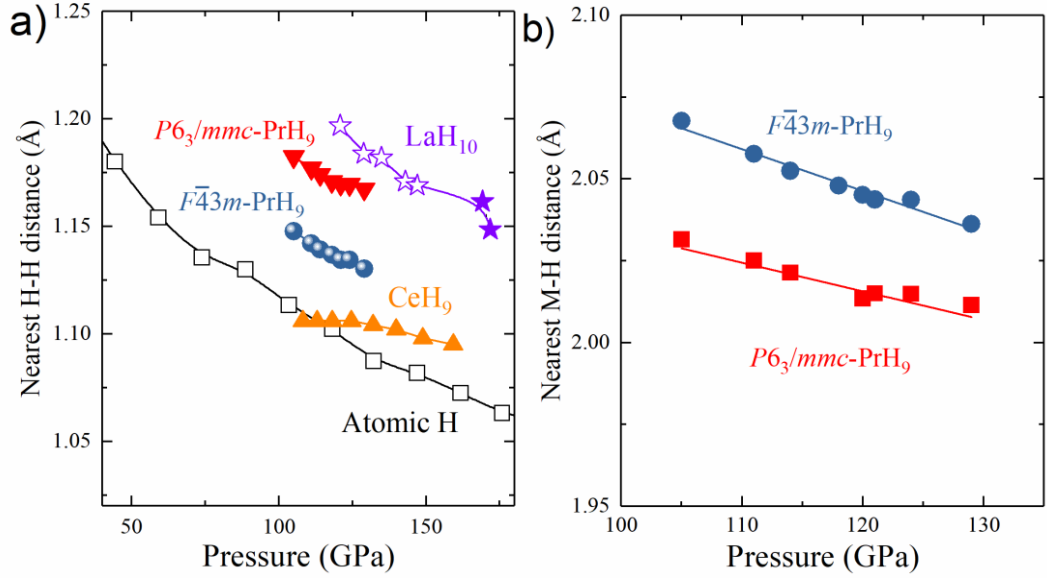

**Fig. S4. Pressure dependence of the nearest H-H distances and Nearest Pr-H distances from experimental cell parameters.** (**a**) Pressure dependence of the nearest H-H distances for two PrH$_9$, LaH$_{10}$ (Ref. 5), CeH$_9$ (Ref. 39), and atomic H (62). (**b**) Nearest Pr-H distances as a function of pressure calculated from experimental cell parameters.

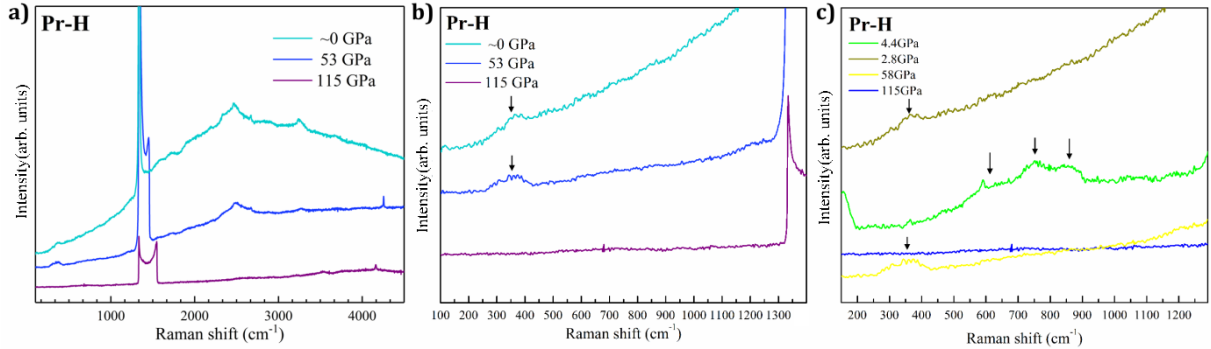

**Fig. S5. Raman spectra of Z1 cell under decompression.** After diamond broken, there was a small peak at 340 cm$^{-1}$, which might response to the non-stoichiometric PrH$_{2+x}$ phases according to the analogy with CeH$_{2+x}$ (63).

# Electron and phonon properties of Pr-H phases

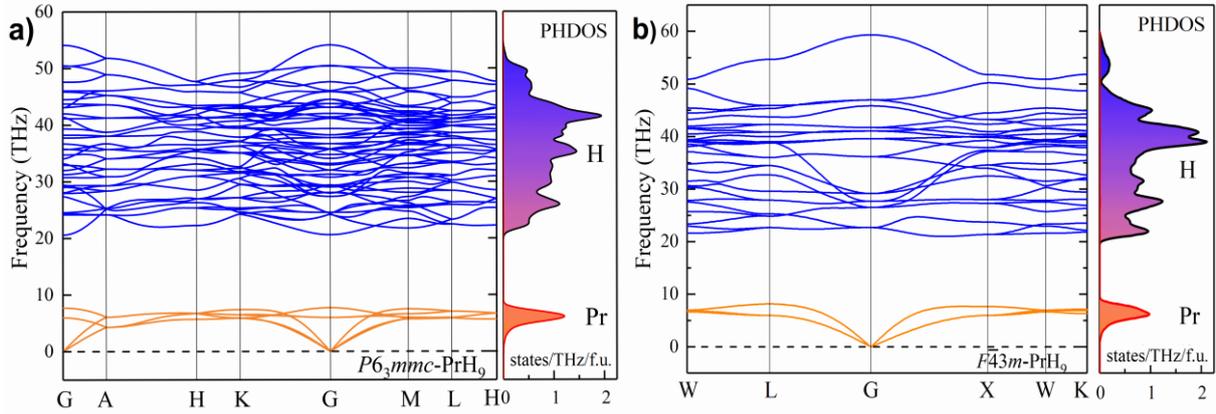

**Fig. S6.** Calculated phonon density of states and band structure for (**a**) $P6_3/mmc$-PrH$_9$ and (**b**) $F\bar{4}3m$-PrH$_9$ at 120 GPa that were used to calculate zero-point energy correction for Pr-H convex hull.

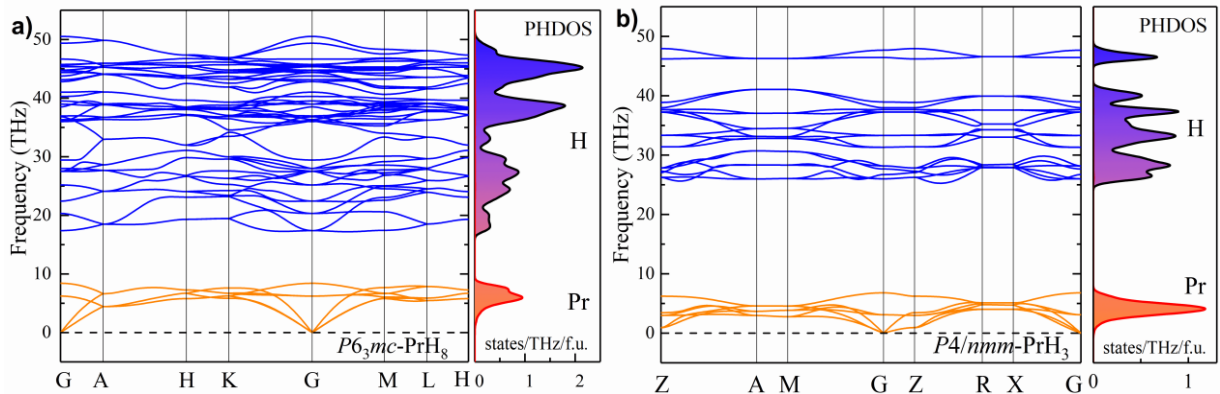

**Fig. S7**. Calculated phonon density of states and band structure for (**a**) $P6_3mc$-PrH$_8$ at 120 GPa and (**b**) $P4/nmm$-PrH$_3$ at 50 GPa.

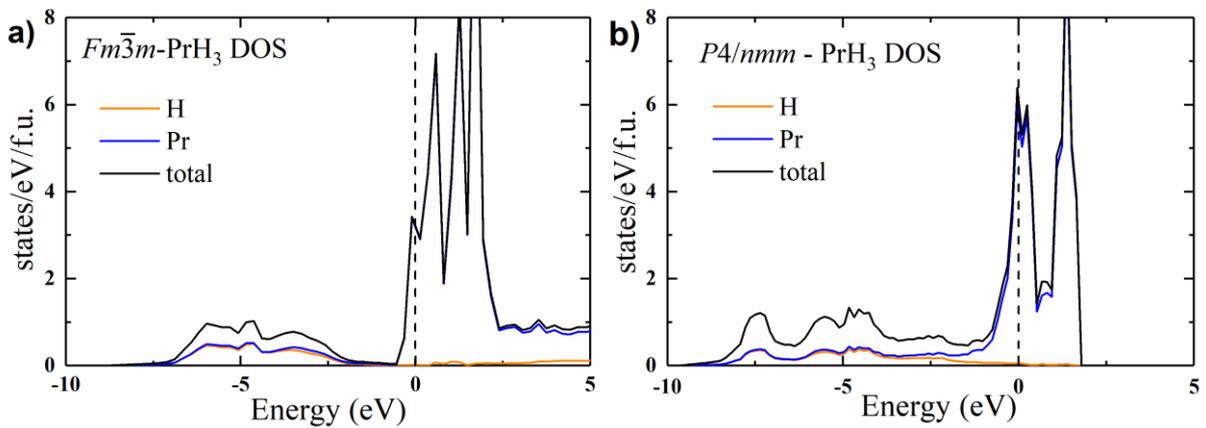

**Fig. S8.** Electron density of states for (a) cubic PrH$_3$ and (b) $P4/nmm$-PrH$_3$ at 30 GPa (per f.u.). DOS($E_F$) caused mostly by Pr *d,f-electrons* and has very high value in both cases that may have negative effect on SC properties.

# Electrical measurements and superconductivity

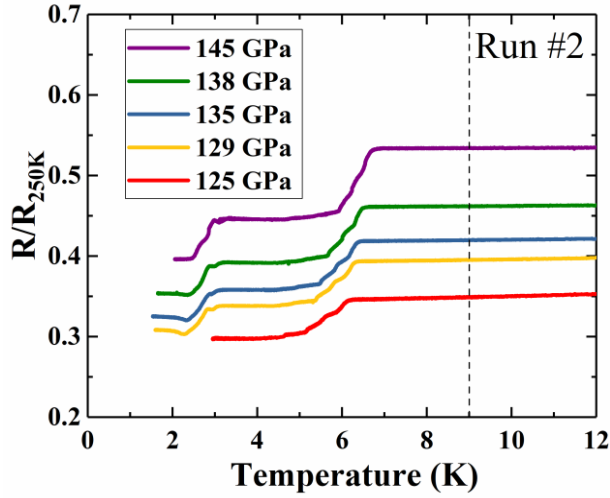

**Fig. S9. Enlarged figure of electrical resistance measurements of PrH$_9$ in sample 2.**

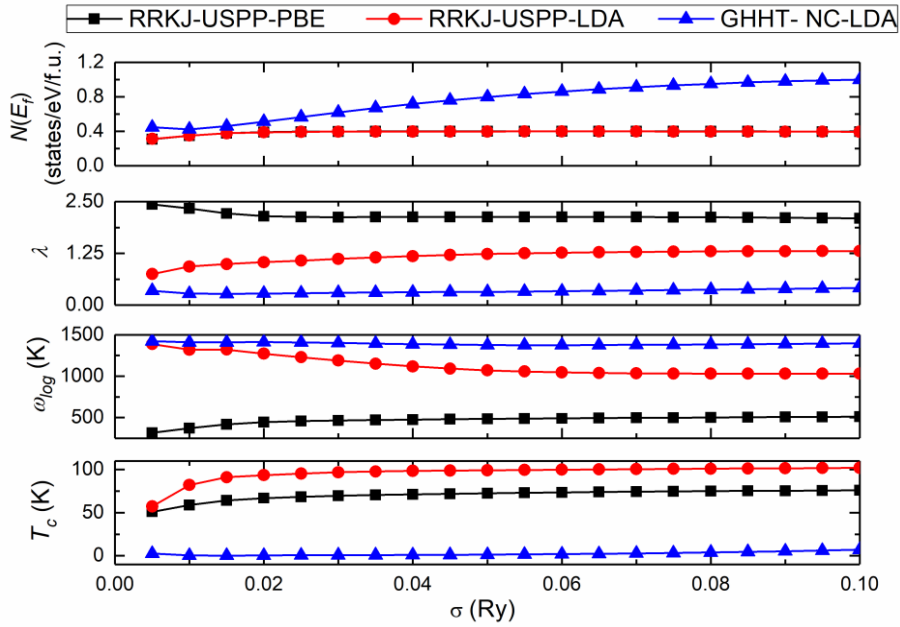

**Fig. S10**. **Calculated superconducting parameters of $F\bar{4}3m$-PrH$_9$ at 120GPa as a function of electronic smearing σ and the pseudopotential.** Density of electronic states N($E_f$) (first panel), EPC coefficient $\lambda$ (second panel), logarithmic frequency $\omega_{log}$ (third panel), and Allen-Dynes modified McMillan critical temperature $T_c$ (fourth panel) of $F\bar{4}3m$-PrH$_9$ at 120GPa as a function of electronic smearing σ and the pseudopotential

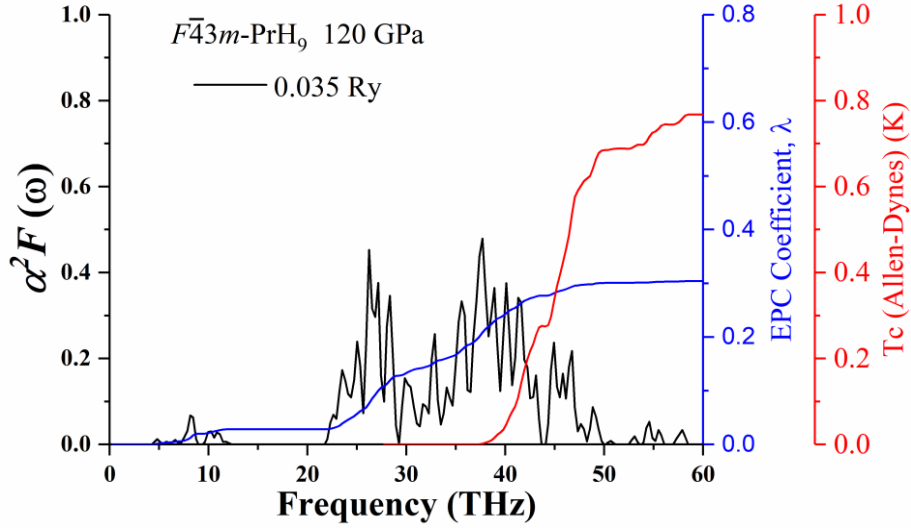

**Fig. S11. Eliashberg spectral functions, the electron-phonon integral $\lambda(\omega)$ and critical transition temperature $T_c(\omega)$ calculated at 120 GPa for cubic PrH$_9$ with $\sigma = 0.035$ Ry.**

We calculated superconducting parameters of cubic PrH$_9$ (120 GPa) with different pseudopotentials (PPs) including the ultrasoft Rappe-Rabe-Kaxiras-Joannopoulos (64, 65) (RRKJ-PBE, RKKL-LDA) and the norm-conserving Goedecker-Hartwigsen-Hutter-Teter (66) (GHHT-NC-LDA) pseudopotentials. The first Brillouin zone was sampled using 6×6×6 q-points mesh, and a denser 24×24×24 k-points mesh was used to calculate the electron wave functions. A kinetic-energy cutoff of 80 Ry for RRKJ and 90 Ry for GHHT was applied. Furthermore, we use Allen-Dynes modified McMillan formula with Coulomb pseudopotential $\mu^* = 0.1$ to estimate $T_c$. The superconducting state computed with ultrasoft RKKJ PPs lead to $T_c$(McM) > 60 K. But the experiment clearly indicated absence of SC transition between 50 and 100 K indicating that ultrasoft RKKJ PPs is not suitable for Pr-H system. Calculations with Norm-Conserving GHHT (LDA) pseudopotential with electronic smearing $\sigma = 0.035$ Ry gave $\lambda < 0.3$ and $T_c < 1$ K., which does not contradict the experimental data.

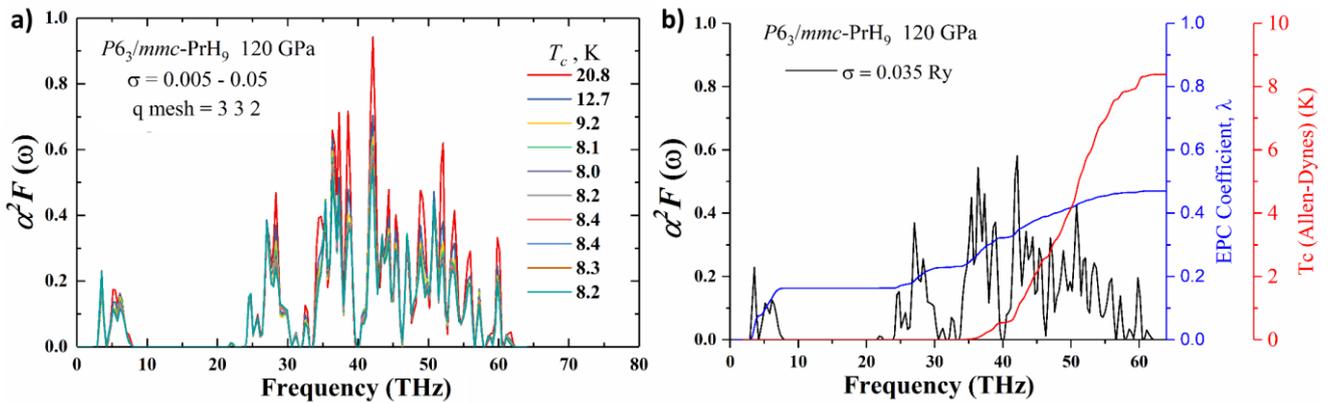

**Fig. S12. A series of Eliashberg spectral functions calculated at 120 GPa for hexagonal PrH$_9$ with (a) q mesh = 3 3 2 and (b) the electron-phonon integral $\lambda(\omega)$ and critical transition temperature $T_c(\omega)$ with $\sigma = 0.035$ Ry.** This diagram illustrates the high sensitivity of the $T_c$ calculations (by Allen-Dynes modified McMillan formula, $\mu^*$=0.1) due to quite complex structure of DOS near the Fermi level.

We continue to calculate superconducting parameters of another superhydrides $P6_3/mmc$-PrH$_9$ using norm-conserving GHHT (LDA) pseudopentiatial, as shown in Fig. S12, where the electron-phonon interaction should be much stronger. A larger $T_c$ of 8.4 K at 120 GPa is obtained, which are in good agreement with the experimental data.